\documentclass[review]{elsarticle}

\usepackage{lineno,hyperref}

\usepackage{amsmath,amssymb}

\usepackage[top=40truemm,bottom=40truemm,left=24truemm,right=24truemm]{geometry}
\usepackage[hang, small,labelfont=bf,up,up]{caption} 
\usepackage{booktabs} 

\usepackage{enumitem} 
\setlist[itemize]{noitemsep} 

\begin{document}

\begin{frontmatter}



\title{%
\begin{flushright}%
{\small ILD-SOFT-2022-001\\KYUSHU-RCAPP-2022-03\\}
\end{flushright}%
{Development of a Vertex Finding Algorithm using Recurrent Neural Network}}

\author[myaddress1]{Kiichi Goto}
\ead{goto@epp.phys.kyushu-u.ac.jp}
\author[myaddress1,myaddress2,myaddress3]{Taikan Suehara}
\ead{suehara@phys.kyushu-u.ac.jp}
\author[myaddress1,myaddress2,myaddress3]{Tamaki Yoshioka}
\author[myaddress4]{Masakazu Kurata}
\author[myaddress5]{Hajime Nagahara}
\author[myaddress5]{Yuta Nakashima}
\author[myaddress5]{Noriko Takemura}
\author[myaddress5,myaddress6,myaddress7,myaddress8]{Masako Iwasaki}

\address[myaddress1]{Department of Physics, Graduate School of Science, Kyushu University}
\address[myaddress2]{Department of Physics, Faculty of Science, Kyushu University}
\address[myaddress3]{Research Center for Advanced Particle Physics (RCAPP), Kyushu University}
\address[myaddress4]{Department of Physics, Graduate School of Science, The University of Tokyo}
\address[myaddress5]{Osaka University Institute for Datability Science (IDS)}
\address[myaddress6]{Department of Mathematics and Physics, Graduate School of Science, Osaka City University}
\address[myaddress7]{Nambu Yoichiro Institute of Theoretical and Experimental Physics (NITEP), Osaka City University}
\address[myaddress8]{Research Center for Nuclear Physics (RCNP), Osaka University}

\date{}
\begin{abstract}
Deep learning is a rapidly-evolving technology with the possibility to significantly improve the physics reach of collider experiments. In this study we developed a novel vertex finding algorithm for future lepton colliders such as the International Linear Collider. We deploy two networks: one consists of simple fully-connected layers to look for vertex seeds from track pairs, and the other is a customized Recurrent Neural Network with an attention mechanism and an encoder-decoder structure to associate tracks to the vertex seeds.
The performance of the vertex finder is compared with the standard ILC vertex reconstruction algorithm.
\end{abstract}


\end{frontmatter}


\section{Introduction}

Machine learning has long been used for event reconstruction and analysis in particle physics.
Deep learning (DL) techniques, which have advanced rapidly in recent years and
are widely applied to various fields of science and technology such as image recognition and automatic translation,
have also started to be applied to particle physics as a natural extension of traditional machine learning~\cite{Albertsson:2018maf, Shlomi:2020ufi, Guest:2018yhq}.
Application of DL methods to the event reconstruction can open new possibilities such as efficiently using parallel computing resources and using various methods on optimization and tuning being developed in DL studies.

The International Linear Collider (ILC)~\cite{Behnke:2013xla} is an $e^+e^-$ collider project being considered for construction in Japan
with an initial center-of-mass energy of 250 GeV. One of the main targets of the ILC is precise measurement of the Higgs boson couplings to various particles, which will provide critical information to search for and identify Beyond-Standard-Model (BSM) theories. Measurements of final states including heavy-flavor ($b$ or $c$) quarks
are especially important in Higgs studies since the Higgs boson couples more strongly to heavier particles.

A major discriminant between $b$ and $c$ quarks from light quarks ($u$/$d$/$s$) is the existence of 
secondary vertices in the jets, since $b$ and $c$ hadrons have finite decay lengths ($c\tau$) of
400-500 $\mu$m and 20-300  $\mu$m, respectively. Since $b$ hadrons mostly decay to $c$ hadrons, $b$ jets usually
have secondary and tertiary vertices corresponding to the decays of $b$ and $c$ hadrons, while $c$ jets have only secondary vertices. 
Secondary and tertiary vertices can be identified by finding points in space at which multiple charged tracks meet within their uncertainties, and which are significantly separated from the event's interaction point (IP) or primary vertex.
The standard method of primary and secondary vertex finding used by ILC experiments is LCFIPlus~\cite{LCFIPlusPaper}, an integrated jet reconstruction tool consisting of vertex finder, jet clustering and jet flavor tagging algorithms.
The secondary vertex finder in LCFIPlus is based on the ``build-up'' technique, which finds track-pairs whose intersection is compatible with secondary vertices as vertex candidates, to which it adds other tracks which pass certain quality selection criteria.
It depends on many human-tuned parameters for the selection.

In this study, we developed a new vertex finding algorithm using a recurrent neural network (RNN) and an attention mechanism~\cite{BahdanauAttention, LuongAttention, AttentionIsAllYouNeed}.
The RNN is a network to process sequential data, often used for speech recognition and natural language processing (NLP).
Attention is an emerging technique in DL, initially developed to improve RNN-based networks.
Our vertex finder is designed to replace the vertex finder in LCFIPlus, facilitating direct comparison of their performance.
Tensorflow (2.1.0)~\cite{TensorflowWeb} and Keras (2.3.1)~\cite{KerasWeb} are used as the  DL framework
for the design and training of the network.
An essential benefit of the ML-based algorithm is to utilize generic methods of optimization ({\it ie}. training and hyperparameter tuning) of the networks for minimum-biased comparison of the performance with various configurations. 
This feature can be used to determine configurations and parameters of the detectors based on consideration of physics performance. Manual selection-based algorithms (as LCFIPlus), in contrast, should be optimized for each configuration with dedicated algorithms, which need intensive work of experts with limited availability and also more easily introduce biases on the detector optimization.

Event samples fully simulated in the International Large Detector (ILD)~\cite{ILD:2020qve}, one of the detector concepts for the ILC, were used to train the vertex finder and to estimate its performance throughout this study.
The ILD detector features impact parameter resolution of a few $\mu$m for high transverse momentum tracks as well as precise tracking detectors and high-granular calorimetry 
optimized for particle flow reconstruction, resulting in excellent heavy-flavor jet identification and reconstruction.
Two-fermion samples of $e^+e^- \to b\bar{b}$ and $c\bar{c}$ at $\sqrt{s}=91$ GeV, produced by the Whizard event generator~\cite{WHIZARDpaper},
and Pythia hadronization~\cite{PYTHIApaper} are used as the input of the detector simulation. Initial-state radiation, finite beam spot size and beam-related background are not considered in this study, since those effects are expected to be small after appropriate removal algorithms as studied in \cite{LCFIPlusPaper} and there are no ILC beam parameters defined at $\sqrt{s}=91$ GeV. 
The jet energy of 2-jet final states in $\sqrt{s} = 91$ GeV largely corresponds to the main target of the Higgs factory of $\sqrt{s} = 250$ GeV, Higgs decays giving 4-6 fermions in the final states, thus the performance of vertex finding is expected to approximate the performance of these final states. 
Details of the usage of the data samples are summarized in Appendix A.



\section{Networks} \label{SNetworks}

The basic concept of our vertex finder is ``build-up'', similar to LCFIPlus.
We use two networks to realize a DL-based build-up vertex finder.\footnote{The paper\cite{Shlomi:2020ufi} also proposes DL-based methods, but they are not two-step algorithms like this paper or LCFIPlus and do not use the attention mechanism.}
The first ``seed-finding'' network considers whether track pairs form a suitable vertex ``seed'' candidate, and the second ``vertex production'' network determines if remaining tracks should be associated to candidate seed vertices.
Detailed structures and hyperparameters of the networks are shown in Appendix B.

\subsection{Network for track pairs - Seed finding} \label{SNetworkforTrackPairs}

The network for track pairs is one of the two networks for vertex finding.
Given as input the parameters listed in Table \ref{TInputVariables OneTrackInformation} of two tracks, the network predicts the vertex's type, as listed in Table \ref{TTypeTheVertexThatTrackPairsOrigin}, and its distance from the IP.
Momentum and charge are duplicated data, which are included to let the network take more notice on them.

\begin{table}[htb]
 \centering
\small
  \begin{tabular}{l l r} \hline
    Name & Description & Number of parameters\\ \hline
    track parameters & $d_0$, $z_0$, $\phi$, $\omega$, $\tan{\lambda}$ &  5\\
    covariance matrix & upper triangle  of the covariance matrix of the track parameters & 15\\
    momentum & magnitude of the momentum of the track & 1\\
    charge & charge of the track & 1\\\hline
  \end{tabular}
  \caption{List of the input variables of a single track. $z_0$ and $d_0$ are the impact parameters along the beam axis and perpendicular to the beam axis, respectively,
  $\phi$ is the initial angle of the track in the $r\phi$ plane, $\omega$ is the signed curvature of the track and $\lambda$ is dip angle.}
  \label{TInputVariables OneTrackInformation}
\end{table}

\begin{table}[htb]
 \centering
  \small
  \begin{tabular}{l l} \hline
    Name & Description\\ \hline
    NC & tracks from different vertices (Not Connected)\\
    PV & track pair from the primary vertex\\
    SVCC & track pair from a secondary vertex of charm flavor in the final state of $c\bar{c}$\\
    SVBB & track pair from a secondary vertex of bottom flavor in the final state of $b\bar{b}$\\
    TVCC & track pair from a tertiary vertex of charm flavor in the final state of $b\bar{b}$\\
    SVBC & one track originated from a $b$ hadron and the other from a $c$ hadron, within the same decay chain\\
    Others & track pair orignated from another particle such as $\tau$, a strange hadron or a photon conversion\\\hline
  \end{tabular}
  \caption{Categorization of track pairs (``seed type'') of the first network.}
  \label{TTypeTheVertexThatTrackPairsOrigin}
\end{table}

The true labels of seed types were obtained from Monte-Carlo (MC) information.
The output of the vertex fitter of LCFIPlus is used as the true label of the vertex distance.

\begin{figure}[htbp]
 \centering
 \begin{minipage}[t]{0.48\hsize}
  \centering
  \raisebox{-30mm}{\includegraphics[trim=100 50 100 100, width= 1.0\textwidth, clip, bb=0 0 1920 1080]{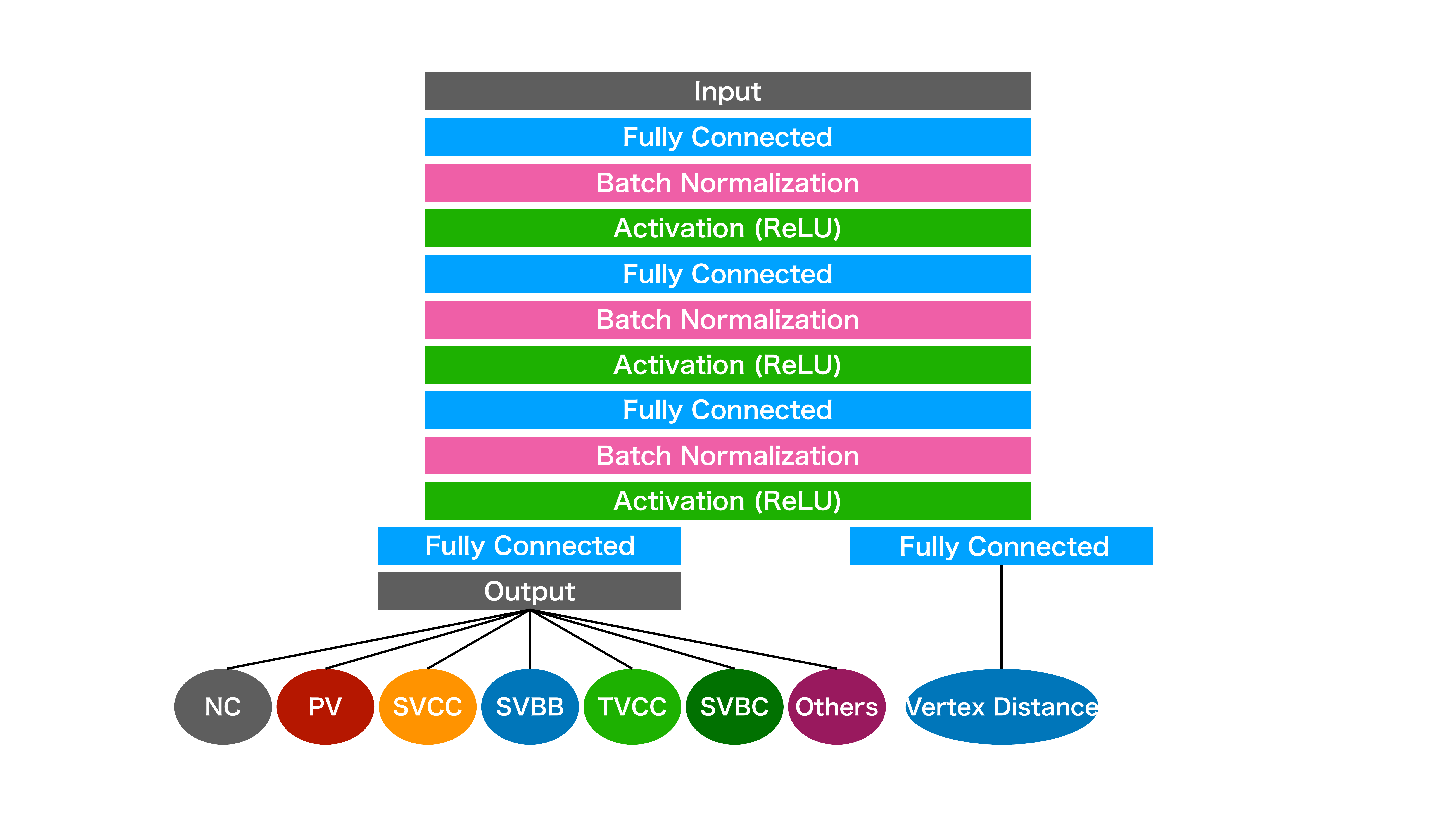}}
  \caption{Structure of the network for track pairs.}
  \label{FNetworkforTrackPairs}
 \end{minipage}
 \begin{minipage}[t]{0.48\hsize}
  \centering
  \raisebox{-30mm}{\includegraphics[trim=50 0 50 0, width=1.0\textwidth, clip, bb=0 0 1920 1080]{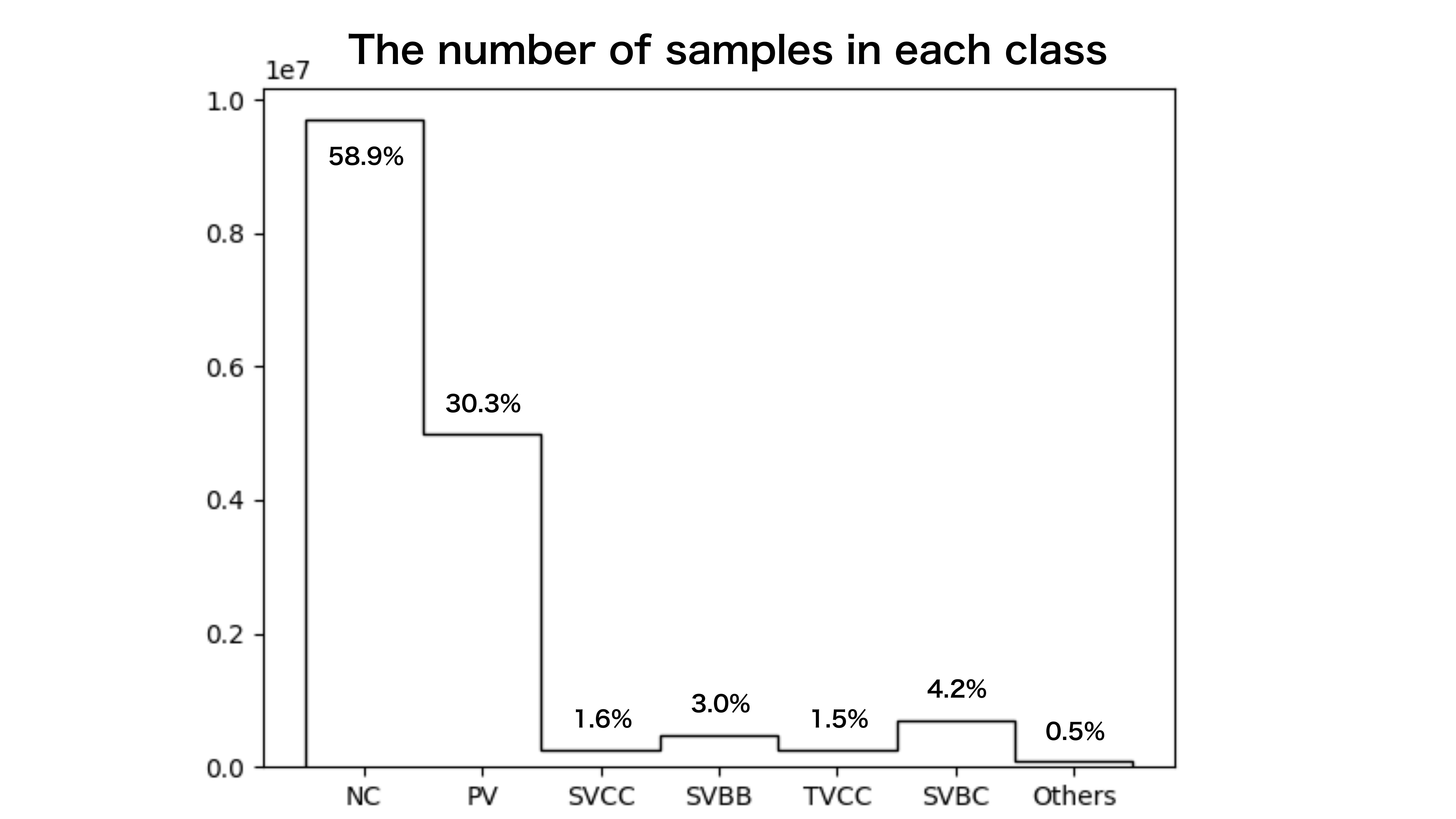}}
  \caption{Statistics of individual categories after the random selection of 50\% in PV and NC categories, for a 1:1 mixture of the $b\bar{b}$ and $c\bar{c}$ samples.}
  \label{FImbalancedData}
 \end{minipage}
\end{figure}

The network processes a track pair, and performs a 7-class categorization of the seed types and a regression to obtain the vertex distance.
The structure is a simple feed-forward network and batch normalization layers are inserted between fully connected layers,
as shown in Figure \ref{FNetworkforTrackPairs}.
The network is divided into two sections, classification and regression, at the last activation layer. 

All track pairs from $b\bar{b}$ and $c\bar{c}$ events (about 147k events each, with 80\% for training and 20\% for validation and 219k events each for test) are used to optimize the network. 
Since the PV and NC categories are provided from both $b\bar{b}$ and $c\bar{c}$ events, these tracks were sampled in half before mixing the events. Figure \ref{FImbalancedData} shows the statistics of each category after the mixing. The Cost-Sensitive Learning scheme~\cite{CostSensitive} is used to weight the loss function accounting for the statistics of the categories to avoid the ineffective training due to the imbalanced statistics.
The total loss function including the regression is as follows:
\begin{equation}
 \begin{split}
&L_{\rm CE} =  - r_{\rm NC}\  t_{\rm NC} \log{(y_{\rm NC})} - r_{\rm PV}\  t_{\rm PV} \log{(y_{\rm PV})} - r_{\rm SVCC}\  t_{\rm SVCC} \log{(y_{\rm SVCC})} - r_{\rm SVBB}\  t_{\rm SVBB} \log{(y_{\rm SVBB})}\\
&- r_{\rm TVCC}\  t_{\rm TVCC} \log{(y_{\rm TVCC})} - r_{\rm SVBC}\  t_{\rm SVBC} \log{(y_{\rm SVBC})} - r_{\rm Others}\  t_{\rm Others} \log{(y_{\rm Others})}\\
&L_{\rm MSLE} =  (\ln{(t_{\rm distance}+1)} - \ln{(y_{\rm distance}+1)})^2\\
&L_{\rm Tot} =  w_{\rm vertex} L_{\rm CE} + w_{\rm distance} L_{\rm MSLE}\\
 \end{split}
\end{equation}
where $L_{\rm CE}$ is the loss function for the 7-class classification, $L_{\rm MSLE}$ is the mean-squared logarithmic error (MSLE) loss function of the vertex distance regression, 
and $L_{\rm Tot}$ is their weighted sum.\\
$r_{\{{\rm NC},\  {\rm PV},\  {\rm SVCC},\  {\rm SVBB},\  {\rm SVBB},\  {\rm TVCC},\  {\rm SVBC},\  {\rm Others}\}}$
are inverse of the number of data in each category to the number of data in the ``Others'' category. 
$t_x$ (= 0 or 1) and $y_x$ are the true labels and predicted scores, respectively.
$w_x$ of the third formula are fixed to mainly train the regression in the early epochs and mainly train the classification in the later epochs.
Providing the vertex distance, an important variable for categorization, helps common fully-connected layers to be more efficiently structured. The vertex distance is used for the vertex selection at later process.

\begin{figure}[htbp]
 \centering
 \includegraphics[trim=0 200 0 200, width=1.0\textwidth, clip, bb=0 0 1920 1080]{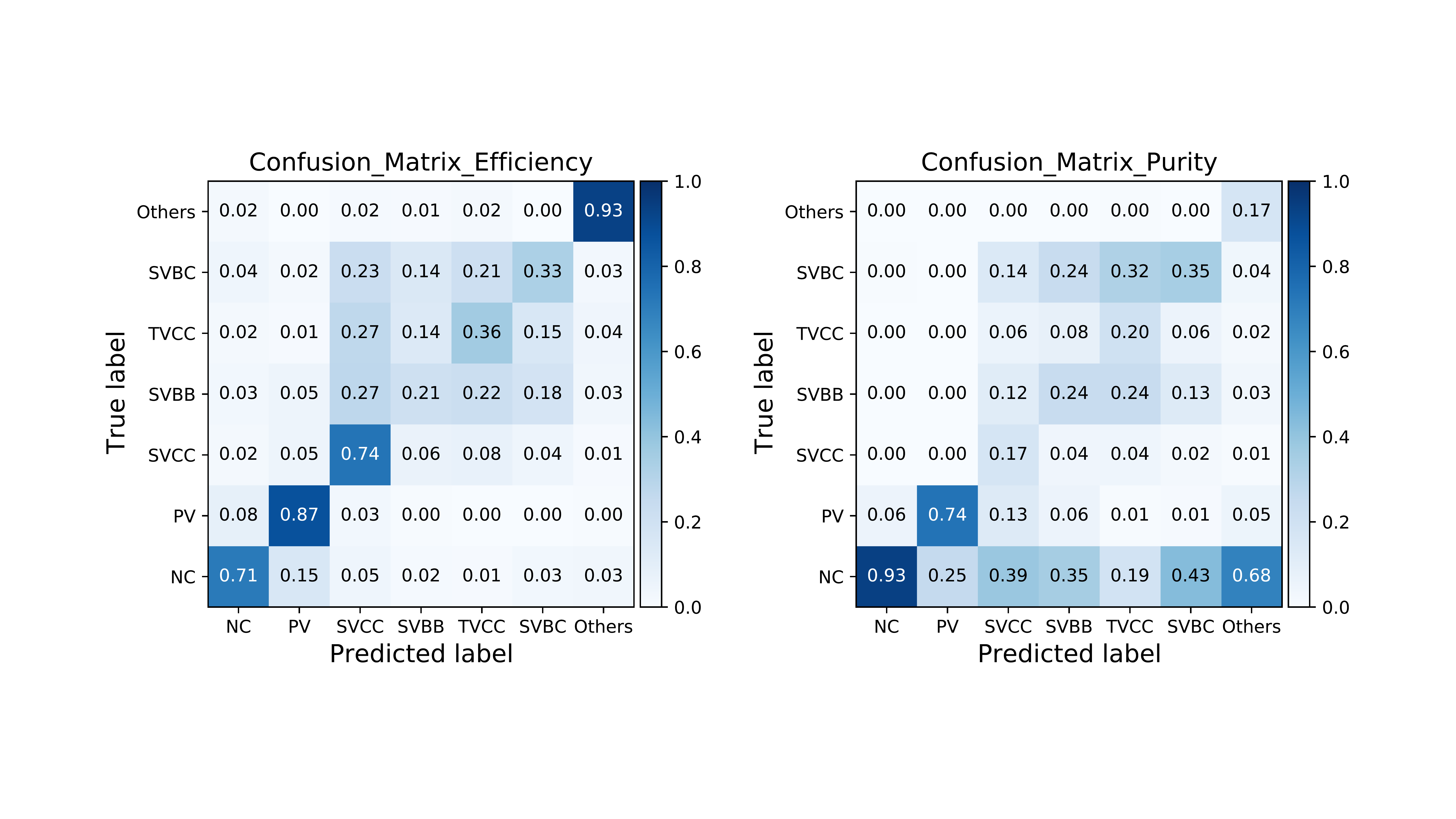}
 \caption{Confusion matrices for efficiency (left) and purity (right). 
 In the efficiency matrix the sum of the each row is unity, while in the purity matrix the sum of the each column is unity. }
 \label{FConfusionMatrix}
\end{figure}

The classification performance of this network is shown in Figure  \ref{FConfusionMatrix}. The efficiencies and purities of NC and PV categories are reasonably high. The vetoing performance that the track pairs from non-SV categories (NC, PV, Others) are not mis-classified to SV categories (SVCC, SVBB, TVCC, SVBC) is 92.3\% and the SV tagging efficiency that the track pairs from SV categories are classified to any SV categories is 90.9\%. Thus a reasonable performance of the classification of NC and PV categories and separation of SV categories from other categories is seen.
While individual SV categories are mixed with other SV categories, all track pairs of these categories are used as secondary vertex candidates in the following process without using detailed categorization and thus the mixture is not harmful. The detailed categorization is aimed to improve the performance by optimizing the network with events of several different natures.
Significant mis-identification of pairs with the true label of NC to other categories is seen in the purity matrix due to the dominant fraction of the true label of NC.
However, since most of the NC-labeled track pairs are composed of a track from primary vertex and one from another vertex, we can reduce the contamination of primary tracks by removing tracks included in the reconstructed primary vertex at the later process.

The weights $r_{\{{\rm NC},\  {\rm PV},\  {\rm SVCC},\  {\rm SVBB},\  {\rm SVBB},\  {\rm TVCC},\  {\rm SVBC},\  {\rm Others}\}}$ are set for equal importance of each category. Increasing weights to categories related to secondary vertices would increase the efficiency to be correctly selected but at a cost of increase of significant contamination from primary and not-connected categories, which is not favorable.


\subsection{Network to associate tracks - Vertex production} \label{SNetworkforAnyTracks}

The second network is used to generate a vertex by adding tracks one by one to each vertex seed obtained by the seed-finding network.
The RNN framework was adopted for this network since the number of tracks varies in each event.
We designed a network to update the vertex using the long short-term memory (LSTM) structure~\cite{LSTMpaper}.
Detailed design of our network is described in the following paragraphs.

\subsubsection*{Standard LSTM Structure}

The standard LSTM contains the following equations.
\begin{equation}
 \begin{split}
  {\mbox{\boldmath{$v$}}}_{N} 
  &= {\mbox{\boldmath{$v$}}}_{N-1} \odot  \sigma (W_f {\mbox{\boldmath{$t$}}}_N + R_f {\mbox{\boldmath{$h$}}}_{N-1}) 
  + \tanh (W_c {\mbox{\boldmath{$t$}}}_N + R_c {\mbox{\boldmath{$h$}}}_{N-1}) \odot  \sigma (W_i {\mbox{\boldmath{$t$}}}_N + R_i {\mbox{\boldmath{$h$}}}_{N-1})\\
  {\mbox{\boldmath{$h$}}}_{N} 
  &= \tanh[{\mbox{\boldmath{$v$}}}_{N-1} \odot \sigma (W_f {\mbox{\boldmath{$t$}}}_N + R_f {\mbox{\boldmath{$h$}}}_{N-1}) 
  + \tanh (W_c {\mbox{\boldmath{$t$}}}_N + R_c {\mbox{\boldmath{$h$}}}_{N-1}) \odot \sigma (W_i {\mbox{\boldmath{$t$}}}_N + R_i {\mbox{\boldmath{$h$}}}_{N-1})] \\
  &\odot \sigma (W_o {\mbox{\boldmath{$t$}}}_N + R_o {\mbox{\boldmath{$h$}}}_{N-1})
 \end{split}
\end{equation}
where $W$, and $R$ are matrices for trainable weights, 
and ${\mbox{\boldmath{$v$}}}_{N}$, ${\mbox{\boldmath{$h$}}}_{N}$ and ${\mbox{\boldmath{$t$}}}_{N}$ are information about the $N^{th}$ hidden states (memory cell), the output and the input, respectively.
The subscripts ``$o$'', ``$f$'', ``$i$'', and ``$c$'' indicate LSTM gates for output, forget, input, and cell, respectively.
``$\odot$'' denotes element-wise multiplication, and $\sigma$ is the sigmoid function. 

Figure \ref{FLSTMStructure} (left) shows a cell of the standard LSTM.
Each cell of the network takes as input a single track and determines whether it should be associated to the vertex whose information is stored in the ``memory''.
If the track is accepted, the ``memory'' is updated with the track.
Parameters of a track pair selected by the seed-finding network
are connected to two fully-connected layers with batch normalization and ReLU activation to calculate the initial state of the memory.

One issue of using the LSTM structure in this network is that LSTM heavily depends on the order in which tracks are provided, while in this application the track order is immaterial. 
On the other hand, we would like to retain the vertex state, which changes with the addition of tracks. 
To reduce the dependence on the order of the tracks, we implemented two extentions as described in the following.

\subsubsection*{Extension 1 --- Dedicated LSTM Structure}

\begin{figure}[htbp]
 \centering
 \includegraphics[trim=0 150 0 200, width=1.0\textwidth, clip, bb=0 0 1920 1080]{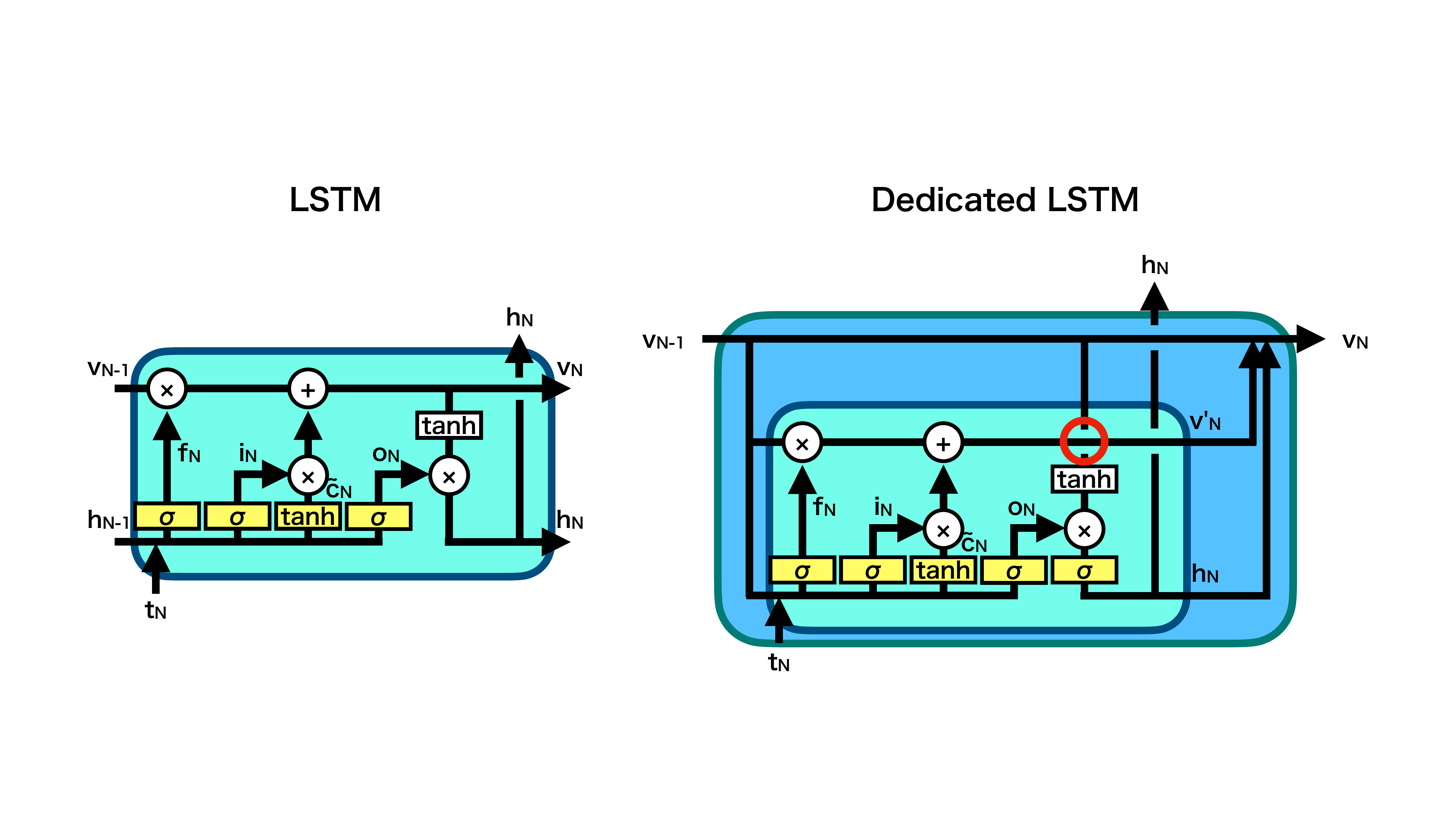}
 \caption{Structure of the standard LSTM (left) and the dedicated LSTM (right). 
 The ``$o_N$'', ``$f_N$'', ``$i_N$'', and ``$\tilde{c}_N$'' indicate LSTM gates for output, forget, input, and cell, respectively. The red circle in right figure means the output is calculated using the old vertex which not containing the Nth track Eq.~(\ref{eq:lstmsteps}). See text for the details.}
 \label{FLSTMStructure}
\end{figure}

Figure \ref{FLSTMStructure} (right) shows a cell of the modified network. 
Each step of the cell (1, 2, 3) in Figure \ref{FDedicatedLSTMCell} is calculated as:
\begin{equation}
\begin{split}
  &1.\ h_{N} 
  = \sigma ({\mbox{\boldmath{$d$}}}_h [\tanh({\mbox{\boldmath{$v$}}}_{N-1}) \odot \sigma (W_o {\mbox{\boldmath{$t$}}}_N + R_o {\mbox{\boldmath{$v$}}}_{N-1}) ])\\
  &2.\ {\mbox{\boldmath{$v'$}}}_{N}
  = {\mbox{\boldmath{$v$}}}_{N-1} \odot \sigma (W_f {\mbox{\boldmath{$t$}}}_N + R_f {\mbox{\boldmath{$v$}}}_{N-1}) 
  + \tanh (W_c {\mbox{\boldmath{$t$}}}_N + R_c {\mbox{\boldmath{$v$}}}_{N-1}) \odot \sigma (W_i {\mbox{\boldmath{$t$}}}_N + R_i {\mbox{\boldmath{$v$}}}_{N-1})\\
  &3.\ {\mbox{\boldmath{$v$}}}_{N} 
  = (1-h_N) {\mbox{\boldmath{$v$}}}_{N-1} + h_N {\mbox{\boldmath{$v'$}}}_{N}
  \end{split}
  \label{eq:lstmsteps}
\end{equation}
where ${\mbox{\boldmath{$d$}}}_h $, $W$, and $R$ are a vector and matrices for trainable weights, 
and ${\mbox{\boldmath{$v$}}}_{N}$ and ${\mbox{\boldmath{$t$}}}_{N}$ are information about the $N^{th}$ hidden states of the vertex and the input track, respectively.
The subscripts ``$o$'', ``$f$'', ``$i$'', and ``$c$'' indicate LSTM gates for output, forget, input, and cell, respectively.
$h_{N}$ is the $N^{th}$ output between 0 and 1, showing whether the $N^{th}$ track is associated to the $(N-1)^{th}$ vertex.
``$\odot$'' denotes element-wise multiplication, and $\sigma$ is the sigmoid function.
The operations of Eq.~(\ref{eq:lstmsteps}) can be understood as follows:

\begin{enumerate}[noitemsep]
 \item determine whether the $N^{th}$ track is associated to the $(N-1)^{th}$ vertex
 \item calculate the updated vertex with the $N^{th}$ track and the $(N-1)^{th}$ vertex
 \item adopt $N^{th}$ vertex if the track is associated in step 1, and keep the $(N-1)^{th}$ vertex if it is not associated
\end{enumerate}

Compared to the original LSTM cell, the hidden state of the short-term memory is effectively removed in the dedicated LSTM cell.
This is expected to reduce the effect of the track ordering and thus provide more robust vertex association regardless of the ordering of the tracks while keeping updating vertex state with associating tracks.

\subsubsection*{Extension 2 --- Attention with Encoder-Decoder Network}

\begin{figure}[htbp]
 \centering
 \begin{minipage}[t]{0.38\hsize}
  \centering
  \raisebox{-30mm}{\includegraphics[width=1.0\textwidth, clip, bb=0 0 1920 1080]{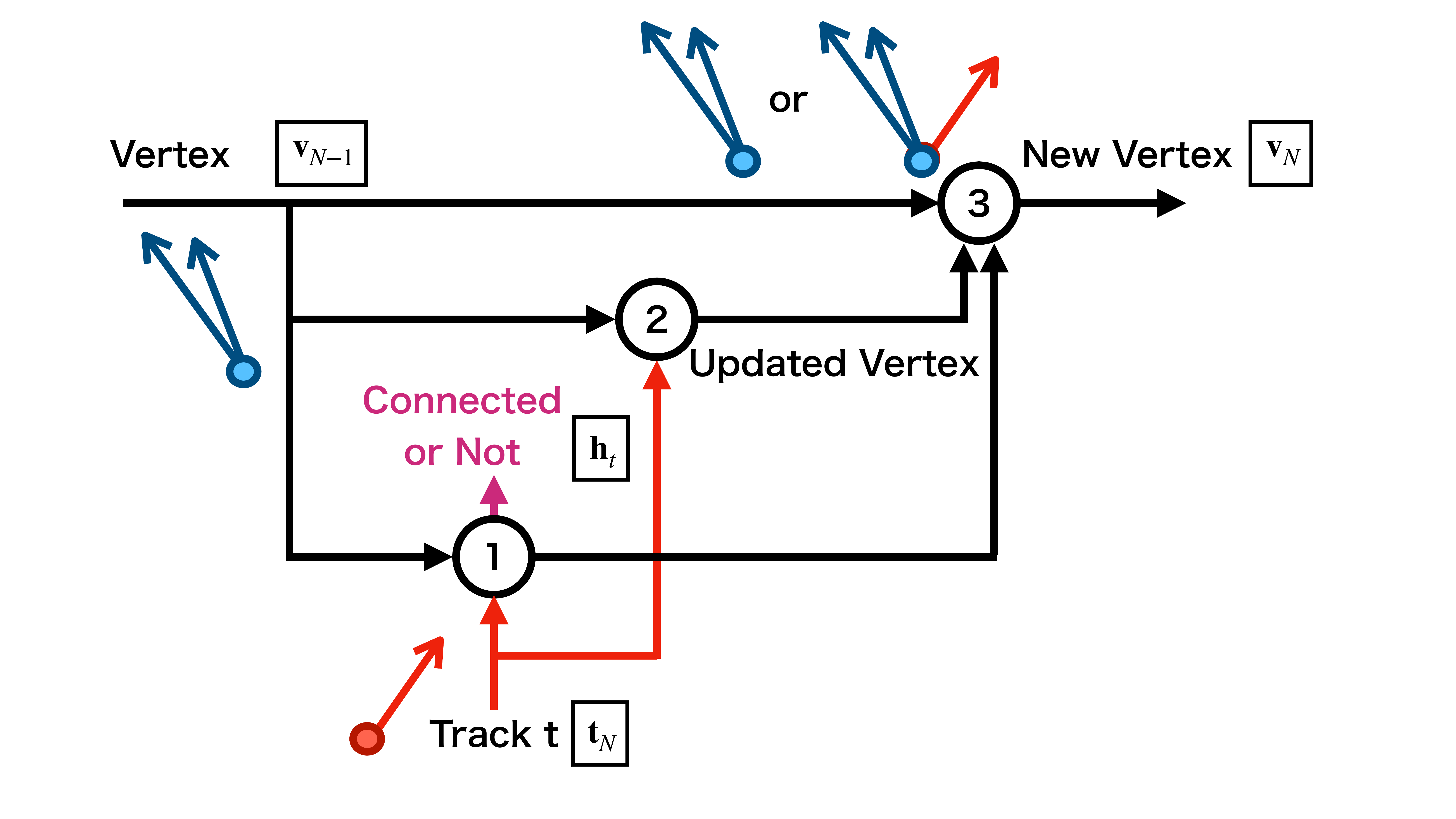}}
  \caption{Schematic of the dedicated LSTM cell. The numbers in the circles stand for the steps formulated in the text. Blue arrows show tracks in the input vertex and red arrows show the input track.}
  \label{FDedicatedLSTMCell}
 \end{minipage}
 \begin{minipage}[t]{0.58\hsize}
  \centering
  \raisebox{-30mm}{\includegraphics[trim=0 200 0 200, width=1.0\textwidth, clip, bb=0 0 1920 1080]{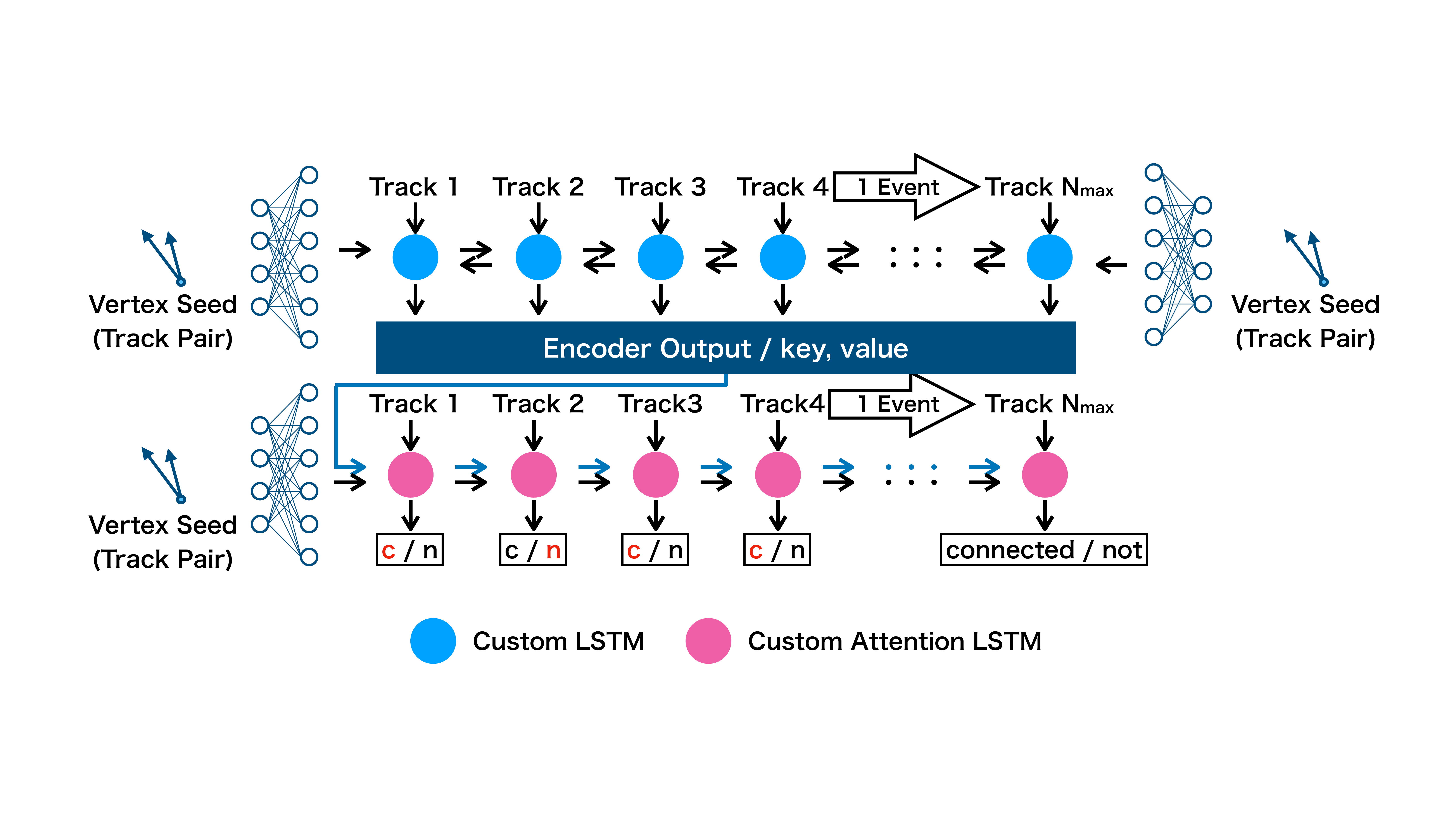}}
  \caption{Schematic of the attention encoder-decoder model. The upper part is the bidirectional LSTM for the encoder, and
  the lower part is the attention LSTM for the decoder. The blue and red circles show custom LSTM cells explained in the text.}
  \label{FEncoderDecoderModel}
 \end{minipage}
\end{figure}

As a further extension, we have implemented an encoder-decoder network with an attention mechanism using a dedicated LSTM cell.
It enables the use of encoded information of all tracks related to the vertex to determine if each track should be associated to the vertex or not with an attention-based decoder.
The attention encoder-decoder model is shown in Figure \ref{FEncoderDecoderModel}.
A bidirectional RNN is used for the encoder part to further reduce the dependence on the order of the tracks.
Encoder (blue) and decoder (red) cells are modified LSTM cells as described above.
The encoder cell is modified as:
\begin{equation}
  {\mbox{\boldmath{$h$}}}_{N} 
  = \tanh({\mbox{\boldmath{$v$}}}_{N-1}) \odot \sigma (W_o {\mbox{\boldmath{$t$}}}_N + R_o {\mbox{\boldmath{$v$}}}_{N-1})
\end{equation}
to provide multi-dimensional variables to the encoder output.

In the decoder cell, attention weights are calculated with the additive attention scheme
using the encoder output as follows:
\begin{equation}
 \begin{split}
{\mbox{\boldmath{$e$}}}_{N} &= {\mbox{\boldmath{$u$}}}_{\rm energy} (K\ U_{\rm key} + T_N\ U_{\rm query})\\
{\mbox{\boldmath{$a$}}}_{N} &= (a_{N,0},\ a_{N,1},\ a_{N,2},\ \cdots a_{N,i},\ \cdots) \\
&= \left(\frac{\exp{({{e}}_{N,0})}}{\sum_j \exp{({{e}}_{N,j})}},\ \frac{\exp{({{e}}_{N,1})}}{\sum_j \exp{({{e}}_{N,j})}},\ \frac{\exp{({{e}}_{N,2})}}{\sum_j \exp{({{e}}_{N,j})}},\  \cdots \frac{\exp{({{e}}_{N,i})}}{\sum_j \exp{({{e}}_{N,j})}},\ \cdots\right)\\
{\mbox{\boldmath{$c$}}}_{N} &= {\mbox{\boldmath{$a$}}}_{N} V\\
h_{N} &= \sigma ({\mbox{\boldmath{$d$}}}_h [\tanh({\mbox{\boldmath{$v$}}}_{N-1}) \odot \sigma (W_o {\mbox{\boldmath{$t$}}}_N + R_o {\mbox{\boldmath{$v$}}}_{N-1} + C_o {\mbox{\boldmath{$c$}}}_{N}) ])\\
{\mbox{\boldmath{$v'$}}}_{N} &= \tanh (W_c {\mbox{\boldmath{$t$}}}_N + R_c {\mbox{\boldmath{$v$}}}_{N-1} + C_c {\mbox{\boldmath{$c$}}}_{N}) \odot \sigma (W_i {\mbox{\boldmath{$t$}}}_N + R_i {\mbox{\boldmath{$v$}}}_{N-1} + C_i {\mbox{\boldmath{$c$}}}_{N})\\
 &+ {\mbox{\boldmath{$v$}}}_{N-1} \odot \sigma (W_f {\mbox{\boldmath{$t$}}}_N + R_f {\mbox{\boldmath{$v$}}}_{N-1} + C_f {\mbox{\boldmath{$c$}}}_{N})\\
{\mbox{\boldmath{$v$}}}_{N} &= (1-h_N) {\mbox{\boldmath{$v$}}}_{N-1} + h_N {\mbox{\boldmath{$v'$}}}_{N}
 \end{split}
\end{equation}
where ${\mbox{\boldmath{$u$}}}$ and $U$ are a list and matrices of the trainable weights for the additive attention.
The key $K$ and value $V$ are the same matrix of the encoder output.
The $N^{th}$ query $T_N$ is a matrix with the $N^{th}$ track stacked.
${\mbox{\boldmath{$e$}}}_{N}$, ${\mbox{\boldmath{$a$}}}_{N}$, and ${\mbox{\boldmath{$c$}}}_{N}$ are the energy, the attention weights, and the context for the $N^{th}$ query, respectively.
$C_x$ are matrices of the trainable weights for the context.
The first three equations calculate the attention.
The last three equations show the extension of the dedicated LSTM structure.
Initial hidden states ${\mbox{\boldmath{$v$}}}_{0}$ of both the bidirectional RNNs of the encoder part and the single RNN of the decoder part are calculated in the same way as for the simple LSTM case, by two fully-connected
layers with input variables of track-pairs.

With the dedicated LSTM cell, we still have effect of the track ordering since the vertex state at the first track only includes vertex seeds while the last track can see additional information from associated tracks. 
This can be improved with adding an encoder-decoder feature to the LSTM network since information of all tracks should be included in the encoded information. Attention is a mechanism to efficiently derive information of encoded data to the individual selections at the decoder.

\subsubsection*{Training}

Samples of track pairs for the initial state and multiple tracks for the sequential input are necessary
for the training. Track pairs coming from the same vertex in the MC information are used
for the initial states and all tracks in the same event are used for the sequential input.
Since the training of the RNN in the Keras framework requires sequential input with fixed length,
dummy tracks with all track parameters set to zero were used for the padding
if the number of tracks was smaller than the fixed length.
To discriminate dummy tracks one additional variable with a flag of whether it is a dummy track
or a real track is added to the sequential input, resulting in 23 variables in total.
The order of the tracks is shuffled at each training epoch
to further reduce the dependence of the training result on the order of the tracks.
About 35k $b\bar{b}$ and $c\bar{c}$ events are used for optimizing the network, with 80\% for training and 20\% for validation.

\begin{figure}[htbp]
 \centering
  \centering
  \raisebox{-30mm}{\includegraphics[trim=150 0 150 0, width=1.0\textwidth, clip, bb=0 0 1920 1080]{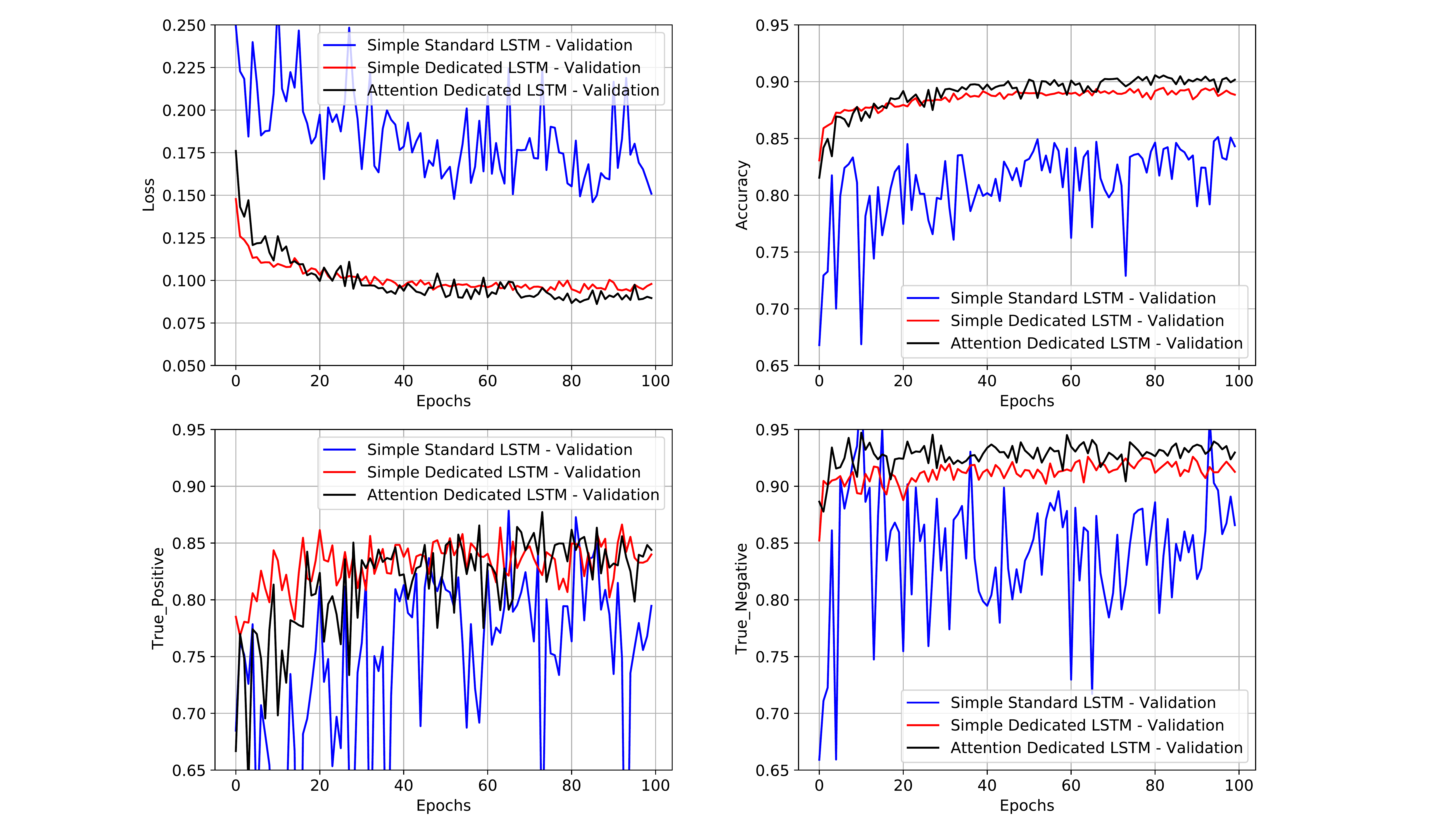}}
  \caption{Comparison of the training curve (upper-left: loss function, upper-right: accuracy, lower-left: true-positive fraction, lower-right: true-negative fraction)
  for the three LSTM structures. See text for the details of each structure.}
  \label{FComparisonwithLSTM}
\end{figure}

Figure \ref{FComparisonwithLSTM} shows the training curves of the three types of networks.
The Simple Standard LSTM stands for the result with the standard LSTM structure,
and the Simple Dedicated LSTM stands for the result with the cell structure of Figure \ref{FDedicatedLSTMCell}
with a simple RNN without the encoder-decoder and attention structures.
The Attention Dedicated LSTM stands for the network described in Figure \ref{FEncoderDecoderModel}.
The two tracks used as the vertex seed were excluded from the calculation of the accuracy, true positive fraction, and true negative fraction.
The clear improvement with the use of the dedicated LSTM structure and the attention encoder-decoder structure
is seen. The instability of the training seen with the standard LSTM may be due to shuffling the track order at each epoch. This effect is not seen in dedicated LSTM models which do not have short-term memories sensitive to track ordering.

\begin{figure}[htbp]
 \centering
  \raisebox{-30mm}{\includegraphics[trim=0 0 0 0, width=0.7\textwidth, clip, bb=0 0 480 270]{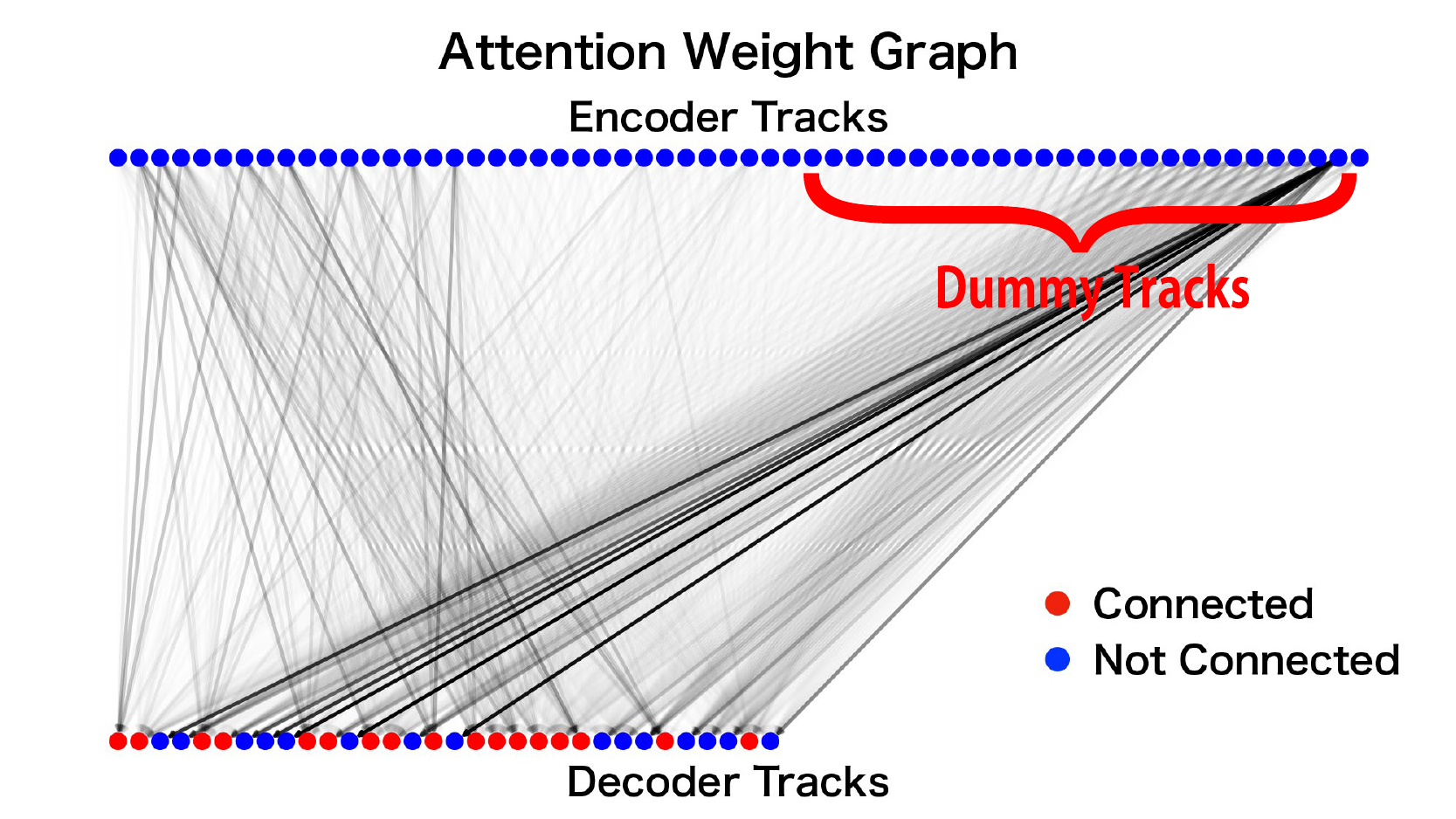}}
  \caption{Strength of attention weights between encoder and decoder tracks of a certain vertex in a single test event.
  Darker lines show higher attention weights. The tracks associated to the vertex are labeled ``Connected'', and the tracks not associated to the vertex are labeled ``Not Connected''.}
  \label{FAttentionWeights}
\end{figure}

Figure \ref{FAttentionWeights} shows the attention weights of one example event independent of training and validation samples.
Each circle shows a track with the same order for the encoder and the decoder tracks.
The difference of the numbers of the tracks is due to the dummy tracks placed only in the encoder tracks.
It shows that since the tracks not associated to the vertex tend to have relatively larger weights to dummy tracks than real tracks, the dummy tracks work for preventing ``not connected'' tracks from receiving information from real tracks.

\section{Vertex Finder with Deep Learning} \label{SVertexFinderwithDeepLearning}

In this section we describe the application of these networks to vertex finding, and compare the resulting performance to an existing algorithm, LCFIPlus.

\subsection{Algorithm}

\begin{figure}[htbp]
 \centering
 \includegraphics[width=0.6\textwidth, clip, bb=0 0 1920 1080]{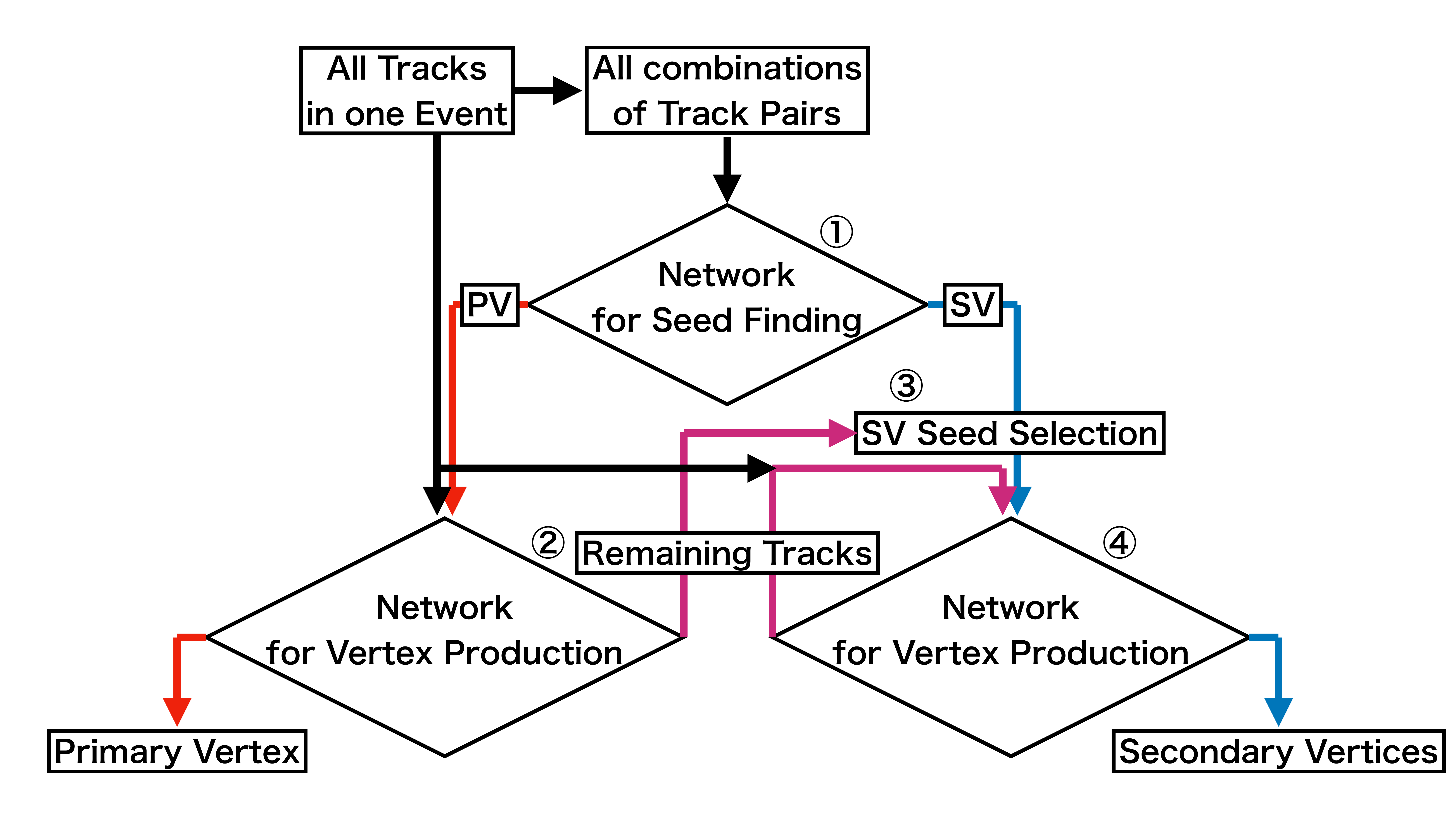}
 \caption{Schematic diagram of our vertex finder. Numbers show corresponding steps in the text. Black arrows are for ``all tracks'' and magenta arrows are for ``remaining tracks''. See text for the details.}
 \label{FAlgorithm}
\end{figure}


Figure \ref{FAlgorithm} shows a schematic diagram of our vertex finder using the two networks.
Primary and secondary vertices are reconstructed by the following steps.

\begin{enumerate}[noitemsep]
 \item Use the ``network for seed finding'' to identify vertex seeds among all pairs of tracks.
 \item Use vertex seeds labeled as PV to generate a primary vertex using the ``network for vertex production''.
 \item Select secondary vertex seeds.
 \item Recurrently generate secondary vertices until selected secondary seeds are exhausted.
\end{enumerate}

In Step 1, every track pair is labeled as PV, SV (SVBB, SVCC, TVCC, SVBC) or others (NC, Others)
by inference of the ``network for seed finding''.
In Step 2, pairs labeled as PV are sorted in descending order of the PV score of the seed-finding network, and
used to calculate the initial state of the ``network for vertex production''.
All tracks in the event are used as sequential input, and tracks with scores larger than a parameter
``score for PV production'' are assigned as tracks from the primary vertex.
This step is repeated according to a parameter ``number of PV seeds'' and all tracks labeled as 
assigned to any of the PV seeds are used to form a combined primary vertex.
Step 3 is a set of preselections of seeds of secondary vertices at thresholds of parameters ``score for SV seeds'' and ``vertex distance''
with the output of the seed-finding network.
The selections are applied to reduce the contamination of NC track pairs misassigned to SV.
Track pairs including track(s) assigned to PV in Step 2 are also removed from the list of seeds.
The remaining track pairs are listed in descending order of the SV score.
Step 4 is the building of secondary vertices with the seeds listed at Step 3.
All tracks in the event are used again as sequential input, and tracks with scores larger than a parameter ``score for SVs production'' are assigned as tracks from the secondary vertex. 
If a track is assigned to both the primary and the secondary vertex, the scores for the primary and secondary vertices are compared. 
When the score of secondary vertex is higher, the track is dropped from the primary vertex, and is assigned to the secondary vertex. 
When the score of primary vertex is higher, the track remains to be assigned to the primary vertex and is not assigned to the secondary vertex. We keep the secondary vertex without the concerned track. Since the seeded tracks of the secondary vertex are kept, we believe that it should not significantly degrade the results of forming secondary vertices.
Step 4 is repeated until all vertex seeds are used. Track pairs including track(s) already assigned to previous secondary vertices
are removed from the list of the seeds and the sequential input.
The parameters written above are summarized with optimized values in Table \ref{TThresholdofTheAlgorithmofTheVertexFinder}.
The ``score for SV seeds'' are total scores of the SVs form ``network for seed finding''.
The ``score for SV seeds'' and ``vertex distance'' are optimized by performance of SV seed selection using Precision-Recall curve.
The threshold of ``vertex distance'' was changed for every 10 (mm) in 10 - 100 (mm).
The other parameters are optimized by performance of vertex finding.
The ``number of PV seed'' were compared with 1-3 seeds and the ``score for PV production'' and the ``score for SVs production'' were tuned by every 0.05 scores. 

\begin{table}[htb]
 \centering
\small
  \begin{tabular}{l l r} \hline
    Name & Description & Value\\ \hline
    score for SV seeds & sum of the scores for SVs obtained by the ``network for seed finding'' & $>0.88$\\
    vertex distance [mm] & distance of the vertex from the IP predicted by the ``network for seed finding'' & $<30.0$\\
    number of PV seeds & number of PV seeds to be used for the initial state of PV production network & 3\\
    score for PV production & score for the PV obtained by the ``network for vertex production'' & $>0.50$\\
    score for SVs production & score for the SVs obtained by the ``network for vertex production'' & $>0.75$\\\hline
  \end{tabular}
  \caption{List of parameters for the vertex finder with optimized values. See text for details of the parameters.}
  \label{TThresholdofTheAlgorithmofTheVertexFinder}
\end{table}

\subsection{Performance} \label{SPerformanceofVertexFinderwithDL}

Table \ref{TPerformance} shows
the combined performance of our vertex finder compared with LCFIPlus,
using about 35k $b\bar{b}$ events  at $\sqrt{s} = 91$ GeV that are statistically independent of the training and validation samples.
Since the event sample and configuration of LCFIPlus is essentially the same as \cite{LCFIPlusPaper}, we assume that LCFIPlus is already optimized in this setup.
In the table each track is categorized according to the MC information as follows.

\begin{itemize}[noitemsep]
 \item Primary: Tracks originating from the primary vertex.
 \item Bottom: Tracks whose parents with non-zero lifetimes include bottom hadrons, except tracks of the Charm category.
 \item Charm: Tracks whose parents with non-zero lifetimes include charm hadrons (except $B_c$ hadrons).
 \item Others: All the other tracks, such as those from $\tau$ decays, strange hadrons, or photon conversions.
\end{itemize}

The table shows the fraction of tracks in each category associated to the reconstructed secondary vertices.
The tracks in the secondary vertices are further categorized by two criteria:

\begin{itemize}[noitemsep]
 \item from same decay chain: all associated tracks come from
 a single decay chain in MC information, descending from the same $b$ hadron.
 \item from same parent particle: all associated tracks come from
 the same most immediate parent particle with a non-zero lifetime.
\end{itemize}

The table shows that the track-based efficiency to be associated to the secondary vertices
is 5-10\% higher with the DL-based vertex finder, compared with LCFIPlus,
with $>$ 95\% of the heavy flavor tracks included in a vertex being attached to a vertex all of whose tracks originate from the same decay chain
as seen by comparing the second and third rows of the table.
Contamination from primary and other tracks is, however, higher for the DL-based vertex finder. This finding indicates that the DL-based vertex finder is reasonably functional. 
The processing time is about 0.27 sec/event with a GPU (NVIDIA TITAN RTX). The processing time of LCFIPlus is about 0.75 sec/event with a CPU (Intel Core i7-4771 3.5GHz using a single core).

The vertex finder in LCFIPlus includes removal of bad-quality tracks and rejection of $V_0$ tracks ($\it ie$. tracks from a $K_0^S$, $\Lambda$ decay or a photon converted at the detector), which are currently not implemented in the current DL-based algorithm. Removing bad-quality tracks and $V_0$ vertices during the seed finding phase will be considered to improve the performance. 
Also we will try to build a new architecture for ``the network for track pairs'' to utilize geometric information in more systematic manners, such as geometric mappings using parametric representation of the tracks. More comprehensive comparison after adding these features will be done as a future work.
Since the optimization of the selection criteria is heavily related to the flavor tagging algorithm, it is also planned to be studied concurrently with the re-implementation of the flavor tagging algorithm with a modern DL-based network. 

\begin{table}[htb]
 \centering
\small
  \begin{tabular}{l l r r r r} \hline
    Algorithm & Track origin & Primary & Bottom & Charm & Others\\ \hline
    & Total number of tracks & 307 657 & 187 283 & 180 143 & 42 888\\\hline
    &  Tracks in secondary vertices & 2.2\% & 63.3\% & 68.4\% & 9.5\%\\
    DL-based & ...from the same decay chain & - & 62.3\% & 67.2\% & -\\
    (this work) & ...from the same parent particle & - & 38.1\% & 36.2\% & 6.4\% \\\hline
    & Tracks in secondary vertices & 0.2\% & 57.9\% & 60.3\% & 0.5\%\\
    LCFIPlus & ...from the same decay chain & - & 57.5\% & 59.9\% & -\\
    & ...from the same parent particle & - & 34.0\% & 37.2\% & 0.3\%\\\hline
  \end{tabular}
  \caption{Performance of the DL-based vertex finder compared with LCFIPlus. See text for explanation of each category.}
  \label{TPerformance}
\end{table}


\section{Summary} \label{SSummary}

A novel vertex finder using DL techniques has been developed.
Two networks were designed: a simple DL architecture with fully-connected layers is used
for the selection of the vertex seeds, and an RNN-based network with a custom cell structure
is used to form vertices by associating tracks to vertex seeds.
An attention mechanism in the encoder-decoder structure has been implemented, and resulted in improved network performance.
The performance of our vertex finder has been compared to LCFIPlus, the standard method used at the ILC, and shows higher efficiency
of the secondary vertex reconstruction at the cost of some increase of the contamination.
We plan to further develop and optimize this algorithm to fully utilize DL techniques in more comprehensive jet analysis including jet clustering and flavor tagging by expanding the network used in this vertex finder.

\section*{Acknowledgements}
The authors would appreciate D.~Jeans and M.~Meyer for useful comments.
We would also like to thank the LCC generator working group and the ILD software working group for providing the simulation and reconstruction tools and producing the Monte Carlo samples used in this study.
This work has benefited from computing services provided by the ILC Virtual Organization, supported by the national resource providers of the EGI Federation and the Open Science GRID.
This work is done in collaboration with the RCNP Project
``Application of deep learning to accelerator experiments''.
Furthermore this work is supported by the U.S.-Japan Science and Technology Cooperation Program in High Energy Physics.

\bibliographystyle{elsarticle-num}
\bibliography{Bibliography}



\appendix

\section{Usage of data samples}

This appendix describes the detailed usage and statistics of data samples
which we used in this study.
Table \ref{DataSamples} lists the samples. $\rm c\bar{c}-$XX shows
samples of charm-pair final states and $\rm b\bar{b}-$XX shows
samples of bottom-pair final states.
$\rm c\bar{c}-$01 and $\rm b\bar{b}-$01 were used for initial investigation
of the characteristics of the samples to design the networks and
also used for making Figure \ref{FImbalancedData}. They were not reused
for further studies to avoid any sample-based bias.
$\rm c\bar{c}$-02 to $\rm c\bar{c}$-05 and $\rm b\bar{b}$-02 to $\rm b\bar{b}$-05
were used for training and validation of two networks. We employed individual samples
for training and validation of two networks not to introduce unexpected correlation
of two networks. Figure \ref{FConfusionMatrix} and \ref{FComparisonwithLSTM}
were produced with the validation samples of each network.
$\rm c\bar{c}$-06 and $\rm b\bar{b}$-06 are the test samples
mainly used for investigating performance, including producing Table \ref{TPerformance}.
Figure \ref{FAttentionWeights} was also made with an event from $\rm b\bar{b}$-06 sample.
These samples are independent and the differences between samples are only random seeds of simulation.

\begin{table}[htbp]
 \centering
 {\fontsize{9pt}{11pt}\selectfont
  \begin{tabular*}{1.0\textwidth}{@{\extracolsep{\fill}}c r r l}\hline
      & the number of events & the number of tracks & usage\\ \hline \hline
    $\rm c\bar{c}-01$ & 70k & 1344k & initial investigation\\ \hline
    $\rm c\bar{c}-02$ & 62k & 1197k & training of ``network for seed finding''\\ \hline
    $\rm c\bar{c}-03$ & 15k & 299k & vaildation of ``network for seed finding''\\ \hline
    $\rm c\bar{c}-04$ & 31k & 598k & training of ``network for vertex production'' \\\hline
    $\rm c\bar{c}-05$ & 8k & 150k & validation of ``network for vertex production'' \\\hline
    $\rm c\bar{c}-06$ & 116k & 2241k & testing both networks including comparison to LCFIPlus\\ \hline\hline
    $\rm b\bar{b}-01$ & 63k & 1326k & initial investigation\\ \hline
    $\rm b\bar{b}-02$ & 56k & 1184k & training of ``network for seed finding''\\  \hline
    $\rm b\bar{b}-03$ & 14k & 296k & validation of ``network for seed finding''\\  \hline
    $\rm b\bar{b}-04$ & 28k & 593k & training of ``network for vertex production''\\ \hline
    $\rm b\bar{b}-05$ & 7k & 148k & vaildation of ``network for vertex production''\\ \hline
    $\rm b\bar{b}-06$ & 103k & 1510k & testing both networks including comparison to LCFIPlus\\ \hline
  \end{tabular*}}
  \caption{The numbers of events and usages of data samples used in this study.}
  \label{DataSamples}
\end{table}

\section{The hyperparameters of the networks}

We show the detailed structures of the networks in Table \ref{TParametersforPairModel} and \ref{TParametersforVLSTMModel}.
The shape of output (the second column) shows the number of nodes in each layer and the ``None'' means batch size.
The previous layers (the third column) shows which the layer is connected to which layers.

Table \ref{TParametersforPairModel} shows the detailed parameters of the ``network for seed finding''.
``Pair Input'', ``Vertex Output'' and ``Position Output''  in layer names (the first column) are the input features of track pairs, the output of 7-class classification for vertex seed and the output of vertex distance, respectively.
``Dense'', ``Batch Normalization'' and ``Activation ReLU'' mean the fully connected layer,  batch normalization and activation function of ReLU, respectively.

Table \ref{TParametersforVLSTMModel} shows the detailed parameters of the ``network for vertex production'' with Attention Dedicated LSTM.
``Pair Input'', ``Encoder Input'' and ``Decoder Input'' in layer names (the first column) are the input features of track pairs, the input features of all tracks in one events with padding to 60 tracks and those without padding, respectively.
The Pair Input are processed the two layer fully connected layer (``Dense'') with activation function of ReLU (``Activation ReLU'').
They are proceeded separately in three same structure (``Encoder Forward''/``Encoder Backward''/``Decoder'').
Two ``Embedding Dense'' layers embed from the features of tracks to embedding vectors.
``Bidirectional Encoder Dedicated LSTM'' and ``Decoder Attention Dedicated LSTM'' are the dedicated recurrent units which are shown in Figure \ref{FLSTMStructure}, the former is extended to bidirectionally and the latter is with attention.
For the ``network for vertex production'' with Simple Standard LSTM and Simple Dedicated LSTM, we only use the decoder part of Table \ref{TParametersforVLSTMModel}.

\begin{table}[htb]
 \centering
 {\fontsize{9pt}{11pt}\selectfont
 \scalebox{0.8}{
  \begin{tabular*}{1.2\textwidth}{@{\extracolsep{\fill}}l c c l}\hline
    layer names & the shape of output & the previous layers\\\hline\hline
    Pair Input & (None, 44) & \\\hline\hline
    Dense 1 & (None, 256) & Pair Input\\\hline
    Batch Normalization 1 & (None, 256) & Dense 1\\\hline
    Activation ReLU 1 & (None, 256) & Batch Normalization 1\\\hline
    Dense 2 & (None, 256) & Activation ReLU 1\\\hline
    Batch Normalization 2 & (None, 256) & Dense 2\\\hline
    Activation ReLU 2 & (None, 256) & Batch Normalization 2\\\hline
    Dense 3 & (None, 256) & Activation ReLU 2\\\hline
    Batch Normalization 3 & (None, 256) & Dense 3\\\hline
    Activation ReLU 3 & (None, 256) & Batch Normalization 3\\\hline\hline
    Vertex Dense & (None, 7) & Activation ReLU 3\\\hline
    Vertex Output & (None, 7) & Vertex Dense\\\hline\hline
    Position Output & (None, 1) & Activation ReLU 3\\\hline\hline
  \end{tabular*}}}
  \caption{The detailed structure of the ``network for seed finding''}
  \label{TParametersforPairModel}
\end{table}

\begin{table}[htb]
 \centering
 {\fontsize{9pt}{11pt}\selectfont
 \scalebox{0.8}{
  \begin{tabular*}{1.2\textwidth}{@{\extracolsep{\fill}}l c c l}\hline
    layer name & the shape of output & the previous layers\\\hline\hline
    Pair Input & (None, 44) &\\\hline
    Encoder Input & (None, 60, 23) &\\\hline
    Decoder Input & (None, None, 23) &\\\hline\hline
    Encoder Forward Dense 1 & (None, 256) & Pair Input\\\hline
    Encoder Backward Dense 1 & (None, 256) & Pair Input\\\hline
    Encoder Forward Activation ReLU 1 & (None, 256) & Encoder Forward Dense 1\\\hline
    Encoder Backward Activation ReLU 1 & (None, 256) & Encoder Backward Dense 1\\\hline
    Encoder Forward Dense 2 & (None, 256) & Encoder Forward Activation ReLU 1\\\hline
    Encoder Backward Dense 2 & (None, 256) & Encoder Backward Activation ReLU 1\\\hline
    Encoder Forward Activation ReLU 2 & (None, 256) & Encoder Forward Dense 2\\\hline
    Encoder Backward Activation ReLU 2 & (None, 256) & Encoder Backward Dense 2\\\hline\hline
    Encoder Embedding Dense & (None, 60, 256) & Encoder Input\\\hline\hline
    Bidirectional Encoder Dedicated LSTM & (None, 60, 512) & Encoder Embedding Dense\\
    &&Encoder Forward Activation ReLU 2\\
    &&Encoder Backward Activation ReLU 2\\\hline
    Reshape Bidirectional Encoder & (None, 27136) & Bidirectional Encoder Dedicated LSTM\\\hline\hline
    Decoder Dense 1 & (None, 256) & Pair Input\\\hline
    Decoder Activation ReLU 1 & (None, 256) & Decoder Forward Dense 1\\\hline
    Decoder Dense 2 & (None, 256) & Decoder Forward Activation ReLU 1\\\hline
    Decoder Activation ReLU 2 & (None, 256) & Decoder Forward Dense 2\\\hline\hline
    Decoder Embedding Dense & (None, None, 256) & Encoder Input\\\hline\hline
    Decoder Attention Dedicated LSTM & (None, None, 1) & Decoder Embedding Dense\\
&& Reshape Bidirectional Encoder\\
&& Decoder Activation ReLU 2\\\hline\hline
  \end{tabular*}}}
  \caption{The detailed structure of the ``network for vertex production''}
  \label{TParametersforVLSTMModel}
\end{table}

\end{document}